\documentclass{elsart}
\usepackage{graphicx}

\begin{document}
\begin{frontmatter}
\title{Estimation Prospects of the Source Number Density of Ultra-high-energy Cosmic Rays}

\author[utap]{Hajime Takami}\footnote{E-mail addresses: takami@utap.phys.s.u-tokyo.ac.jp (H.Takami), sato@phys.s.u-tokyo.ac.jp (K.Sato)} and 
\author[utap,resceu]{Katsuhiko Sato}

\address[utap]{Department of Physics, School of Science, the University of Tokyo, 7-3-1 Hongo, Bunkyo-ku, Tokyo 113-0033, Japan}
\address[resceu]{Research Center for the Early Universe, School of Science, the University of Tokyo, 7-3-1 Hongo, Bunkyo-ku, Tokyo 113-0033, Japan}

\begin{abstract}
We discuss the possibility of accurately estimating the source number density 
of ultra-high-energy cosmic rays (UHECRs) using small-scale anisotropy 
in their arrival distribution. 
The arrival distribution has information on their source and 
source distribution. 
We calculate the propagation of UHE protons in a structured 
extragalactic magnetic field (EGMF) and simulate 
their arrival distribution at the Earth using our previously developed method. 
The source number density that can best reproduce observational 
results by Akeno Giant Air Shower Array is estimated 
at about $10^{-5}~{\rm Mpc}^{-3}$ in a simple source model. 
Despite having large uncertainties of about one order of magnitude,  
due to small number of observed events in current status, 
we find that more detection of UHECRs in the Auger era 
can sufficiently decrease this so that 
the source number density can be more robustly estimated. 
200 event observation above $4 \times 10^{19}~{\rm eV}$ 
in a hemisphere can discriminate 
between $10^{-5}$ and $10^{-6}~{\rm Mpc}^{-3}$. 
Number of events to discriminate between $10^{-4}$ 
and $10^{-5}~{\rm Mpc}^{-3}$ 
is dependent on EGMF strength. 
We also discuss the same in another source model in this paper. 
\end{abstract}


\end{frontmatter}

%%%%%%%%%%%%%%%%%%%%%%%%%%%%%%%%%%
%%%%%%%%%%%%%%%%%%%%%%%%%%%%%%%%%%
\section{Introduction} \label{introduction}
%%%%%%%%%%%%%%%%%%%%%%%%%%%%%%%%%%
%%%%%%%%%%%%%%%%%%%%%%%%%%%%%%%%%%

The nature of the sources of ultra-high-energy cosmic rays (UHECRs) 
is still poorly known 
though many efforts in detection of these particles with UHE energies. 
It is one of challenging problems in astroparticle physics. 

About 10 years ago, 
Akeno Giant Air Shower Array (AGASA) reported small-scale anisotropy 
of the UHECR arrival distribution within its angular resolution 
while it did large-scale isotropy with a harmonic analysis 
\cite{takeda99,takeda01}. 
It found five doublets and one triplet as event clusterings. 
It is enough evidence that 
origin of UHECRs is astrophysically point-like source. 
However, no obvious astronomical 
counterparts to the observed UHECRs have been found. 
One of this reasons is magnetic fields in the universe. 

Extragalactic magnetic field (EGMF) is poorly known 
theoretically and observationally. 
As an upper limit for its strength, $B$, 
and correlation length, $l_c$, 
$B~{l_c}^{1/2} < (1~{\rm nG})(1~{\rm Mpc})^{1/2}$, 
by Faraday rotation measurements of radio signals 
from distant quasars, is often adopted \cite{kronberg94}. 
Based on its upper limit, 
UHECR propagation has been discussed 
in simply uniform turbulent magnetic field 
\cite{yoshiguchi03,aloisio04,berezinsky06}. 

It is also observationally known 
that clusters of galaxies have strong magnetic fields 
with 0.1 - a few $\mu$G at its center \cite{vallee04}. 
This is shown that there are not only uniform EGMF but also 
relatively strong EGMF incidental to the local structure. 
Recently, several groups have performed 
simulations of large-scale structure formation 
with magnetic fields \cite{sigl03,sigl04,dolag05}. 
They find that magnetic field traces the local density distribution. 
Such magnetic field plays an important role in UHECR propagation 
since it is stronger than the uniform one.

In our previous study \cite{takami06},  
we constructed a structured EGMF model that reflects the local structures 
and discussed predicted arrival distribution of UHE protons at the Earth 
in consideration with their propagation processes. 
We also considered Galactic magnetic field (GMF) in the propagation. 
As a result, the number density of UHECR sources was constrained 
to $\sim 10^{-4}~{\rm Mpc}^{-3}$ 
under our luminosity-weighted source model 
(explained in section \ref{model}) from observed arrival distribution. 
The EGMF strength was normalized only to $0.4~\mu{\rm G}$ 
at the center of the Virgo cluster. 
However, observational measurements of magnetic field in a galaxy cluster 
result in 0.1 - a few $\mu$G, so that the strength 
of magnetic field has large uncertainty 
of about one order of magnitude. 
Thus, it is important that the propagation process and 
the arrival distribution are investigated 
in several strengths of EGMF. 

In this study, 
we calculate propagation of UHE protons in a structured EGMF 
with several strengths for normalization 
and simulate the arrival distributions at the Earth. 
At first, comparing them to observational results, 
we constrain the source number density. 
We use AGASA data for this purpose 
because AGASA has observed the most number of events. 
Such constraint is dependent on UHECR source model. 
We discuss about two simple source models. 
We find that the number density has large uncertainty 
of about an order of magnitude due to small number of observed events 
at present. 
So, we also discuss the possibility of a decrease 
in the uncertainty with future observations. 

In section \ref{model}, our models of UHECR source distribution, 
structured EGMF, and GMF are briefly explained. 
In section \ref{method}, a numerical method of UHE proton propagation, 
construction of the arrival distribution and 
statistical analysis are explained. 
We present our results in section \ref{result}, 
and summarize in section \ref{summary}.

%%%%%%%%%%%%%%%%%%%%%%%%%%%%%%%%%%
%%%%%%%%%%%%%%%%%%%%%%%%%%%%%%%%%%
\section{Our Models} \label{model}
%%%%%%%%%%%%%%%%%%%%%%%%%%%%%%%%%%
%%%%%%%%%%%%%%%%%%%%%%%%%%%%%%%%%%

In this section, 
our distribution models of UHECR sources 
and magnetic fields are briefly explained. 
These models are almost the same as those in our previous work, 
thus, more detailed explanation is written in reference \cite{takami06}. 

Our models of the source distribution and a structured EGMF are 
constructed out of the {\it Infrared Astronomical Satellite} 
Point Source Catalogue Redshift Survey ({\it IRAS} PSCz) catalog 
of galaxies \cite{saunders00}. 
This catalog is thought to be the most appropriate galaxy catalog 
for these purposes since it has the largest sky coverage 
($\sim 84 \%$ of all the sky). 
The selection effects in the observation are corrected 
with a luminosity function of the {\it IRAS} galaxies \cite{takeuchi03}. 
A set of galaxies after this correction is called 
{\it our sample galaxies} below. 
Our sample galaxies within 100 Mpc are used for construction of our models 
since small number of galaxies can be detected outside 100 Mpc. 
We adopt $\Omega_m = 0.3,~\Omega_{\Lambda} = 0.7$ 
and $H_0 = 71~{\rm km~s^{-1}~{Mpc^{-1}}}$ as the cosmological parameters.

We assume a subset of our sample galaxies to be an UHECR source distribution. 
The number density of UHECR sources is taken as our model parameter. 
For a given source number density, 
we randomly select galaxies from our sample galaxies 
with probabilities proportional to the absolute luminosity of each galaxy. 
Source distribution outside 100 Mpc is assumed to be isotropic and 
luminosity distribution of the galaxies follows the luminosity function. 

In this study, we adopt two simple source models. 
One is a source distribution 
that all sources have the same power for injection 
of UHE protons, called {\it normal source model}, 
and the other is that each source has a power 
proportional to its luminosity, called {\it luminosity-weighted source model}. 
In both the models, it is assumed that 
all sources have the same of maximum acceleration energy. 
Our sample galaxies are over 
about five orders of magnitude in the luminosity. 
We construct 100 source distributions in each source number density.

Several simulations of the large-scale structure formation with magnetic field 
have found that the EGMF roughly traces distribution of the baryon density 
\cite{sigl03,dolag05}. 
According to these results, 
our structured EGMF model is constructed based on simple assumptions. 
We constructed the matter density distribution from our sample galaxies 
with a spatial resolution of $1 \rm{Mpc}$. 
In each cell, 
a strength of the EGMF is related to the matter density, $\rho$, as 
$\vert B_{\rm EGMF} \vert \propto {\rho}^{2/3}$ and 
a turbulent magnetic field with the Kolmogorov spectrum is assumed. 
The strength of EGMF is normalized at a cell 
that contains the center of the Virgo Cluster 
to $0.0, 0.1, 0.4$ and $1.0 \mu{G}$. 
Note that magnetized space covers about 5\% 
of the universe within 100 Mpc. 
The remaining 95\% space has no magnetic field. 
Outside of 100 Mpc, we assume an uniform turbulent field 
with 1 nG. 

As GMF model, 
a model in reference \cite{alvarez02} is adopted. 
This model consists of a bisymmetric spiral field and a dipole field. 
We neglect a turbulent component of the GMF 
though its field strength is in the range of 
$0.5 B_{\rm reg} - 2 B_{\rm reg}$ where $B_{\rm reg}$ is 
the strength of the regular component of the GMF. 
The turbulent component little changes the arrival direction 
of UHE protons \cite{yoshiguchi04a}.

%%%%%%%%%%%%%%%%%%%%%%%%%%%%%%%%%%%%%%%%%
%%%%%%%%%%%%%%%%%%%%%%%%%%%%%%%%%%%%%%%%%
\section{Numerical Methods} \label{method}
%%%%%%%%%%%%%%%%%%%%%%%%%%%%%%%%%%%%%%%%%
%%%%%%%%%%%%%%%%%%%%%%%%%%%%%%%%%%%%%%%%%

%%%%%%%%%%%%%%%%%%%%%
\subsection{Calculation of the Arrival Distribution of UHE Protons} 
\label{method_prop}
%%%%%%%%%%%%%%%%%%%%%

In this study, 
propagation of UHE protons is calculated 
by an application for the backtracking method. 
UHE protons injected from their sources cannot always reach the Earth 
because of their deflections by magnetic fields. 
It wastes many CPU time to calculate the propagation of such protons 
and to construct the arrival distribution. 
In order to solve this problem, 
we suggested a new method of calculation of the propagation 
in intergalactic space 
with attention to an inverse process of UHECR propagation \cite{takami06}. 
In this method, 
we consider UHECRs with the charge of -1 that ejected from the Earth, 
and calculate their trajectories in Galactic space 
and intergalactic space. 
Trajectories of these particles can be regarded 
as trajectories of UHE protons from extragalactic space. 

In intergalactic space, 
we consider not only the deflections due to EGMF, 
but also the energy-loss processes, which are 
adiabatic energy loss due to the expanding universe, 
the electron-positron pair creation and photopion production 
in collision with the Cosmic Microwave Background (CMB) 
\cite{berezinsky88,yoshida93}. 
All of these processes are treated as continuous processes. 

The adiabatic energy loss results from the expansion of the universe. 
This energy loss rate is written as 
\begin{equation}
\frac{d E}{dt} = -\frac{\dot{a}}{a}E = 
-H_0 \left[ \Omega_m (1 + z)^3 + \Omega_{\Lambda} \right]^{1/2} E. 
\end{equation}
For the pair creation 
we adopt the analytical fitting functions 
given by \cite{chodorowski92} to calculate 
the energy-loss rate on isotropic photons. 
For the photopion production, 
the energy-loss length 
which is calculated by simulating the photopion production 
with the event generator SOPHIA \cite{mucke00} is used. 

In Galactic space, 
all energy loss processes are neglected 
since the path lengths of UHE protons are 
enough shorter than energy-loss lengths of those processes.

The arrival distribution of UHE protons at the Earth can 
be generated from their trajectories calculated with a method 
explained above. 
We eject 2,500,000 particles from the Earth isotropically 
with the ejection spectrum of $d N/dE \propto E^{-1.0}$ 
and record their trajectories. 
The trajectories are calculated 
until their propagation times reach the age of the universe 
or their energies reach $10^{22} \rm{eV}$, 
which is assumed as 
maximum energy of accelerated proton at each source. 
With our source models, 
we calculate a factor for each trajectory, 
which represents the relative probability that 
the $j$ th proton reaches the Earth, 
\begin{equation}
P_{\rm selec}(E,j) \propto \sum_i 
\frac{L_{i,j}}{( 1 + z_{i,j}) {d_{i,j}}^2} 
\frac{dN / dE_g (d_{i,j}, {E_g}^{i})}{E^{-1.0}} 
\frac{d E_g}{dE}. 
\end{equation}
Here $i$ labels sources on each trajectory, 
while $z_{i,j}$, $d_{i,j}$, and $L_{i,j}$ are 
the redshift, distance, and luminosity of such sources. 
${E_g}^i$ is the energy of a proton at a source, 
which has the energy $E$ at the Earth. 
Thus, $dN / dE_g (d_{i,j}, {E_g}^i)$ is the energy spectrum 
of UHE protons ejected from a source 
whose distance is $d_{i,j}$. 
We assume that UHE protons are injected with a power-law 
spectrum in the energy range of $10^{19} - 10^{22} \rm{eV}$. 
$d E_g / dE$ represents a correction factor for the variation of the shape of 
the energy spectrum through the propagation. 
The power-law index of $dN / dE_g$ is assumed to be 2.6 
in order to fit the calculated energy spectrum above $10^{19}$eV 
to the observational result \cite{marco03}. 
This corresponds to an injection model 
that the ankle is assumed to be the pair creation dip \cite{berezinsky05}. 

We randomly select trajectories according to these relative
probabilities, $P_{\rm selec}$,  
so that the number of the selected trajectories 
is equal to the required event number. 
The mapping of the ejected direction of each particle at the Earth 
is the arrival distribution of UHE protons. 
If we have to select the same trajectory more than once 
to adjust the number of selected trajectories, 
we generate a new event 
whose arrival direction is calculated by adding a normally 
distributed deviation with zero mean and 
variance equal to the experimental resolution 
to the original arrival direction. 

%%%%%%%%%%%%%%%%%%%%%
\subsection{Statistical Methods} \label{method_stat}
%%%%%%%%%%%%%%%%%%%%%

In this subsection, 
we introduce two statistical quantities needed to compare 
the arrival distribution of UHE protons with that obtained 
by the observation statistically. 

The two-point correlation function $N(\theta)$ contains information 
on the small-scale anisotropy. 
We start from a set of events generated from our simulation. 
For each event, we divide the sphere into concentric bins of 
angular size $\Delta \theta$ and count the number of events 
falling into each bin. 
We then divide it by the solid angle of the corresponding bin, 
that is, 
\begin{equation}
N(\theta) = \frac{1}{2\pi \vert \cos \theta - \cos (\theta + \Delta \theta) 
\vert} \sum_{\theta \leq \phi \leq \theta + \Delta \theta} 
1 ~ \rm{[sr^{-1}]}, 
\end{equation}
where $\phi$ denotes the separation angle of the two events. 
The angle $\Delta \theta$ is taken to be $1^{\circ}$ in this analysis. 
The AGASA data show a correlation at small angles $(\sim 3^{\circ})$ 
with a $2.3 (4.6) \sigma$ significance of deviation 
from an isotropic distribution for 
$E > 10^{19} \rm{eV} ( E > 4 \times 10^{19} \rm{eV} )$ 
\cite{takeda01}.

In order to investigate what number density of sources can best reproduce 
the two-point correlation observed by AGASA, 
we introduce $\chi_{\theta_{\rm{max}}}$ for a source distribution as 
\begin{equation}
\chi_{\theta_{\rm{max}}} = \frac{1}{\theta_{\rm{max}}} 
\sqrt{ \sum_{\theta=0}^{\theta_{\rm{max}}} 
\frac{ \left[ N(\theta) - N_{\rm{obs}}(\theta) \right]^2 }
{{\sigma(\theta)}^2}}, 
\label{chi10}
\end{equation}
where $N(\theta)$ is the two-point correlation function calculated 
from a simulated arrival distribution 
within $-10^{\circ} \leq \delta \leq 80^{\circ}$, 
which corresponds to the sky observed by AGASA and 
$N_{\rm{obs}}(\theta)$ is that obtained from the AGASA data 
at angle $\theta$. 
The $\sigma(\theta)$ is the statistical error of $N(\theta)$ 
due to the finite number of simulated events. 
The random event selection is performed 20 times. 
This $\chi_{\theta_{\rm{max}}}$ represents the goodness of fit 
between the simulated two-point correlation and the observed one. 
In this study, we take $\theta_{\rm{max}}$ to be $10^{\circ}$.

%%%%%%%%%%%%%%%%%%%%%%%%%%%%%%%%%%%%%%%%%
%%%%%%%%%%%%%%%%%%%%%%%%%%%%%%%%%%%%%%%%%
\section{Results} \label{result}
%%%%%%%%%%%%%%%%%%%%%%%%%%%%%%%%%%%%%%%%%
%%%%%%%%%%%%%%%%%%%%%%%%%%%%%%%%%%%%%%%%%

\begin{figure}[t]
\begin{center}
\includegraphics[width=0.48\linewidth]{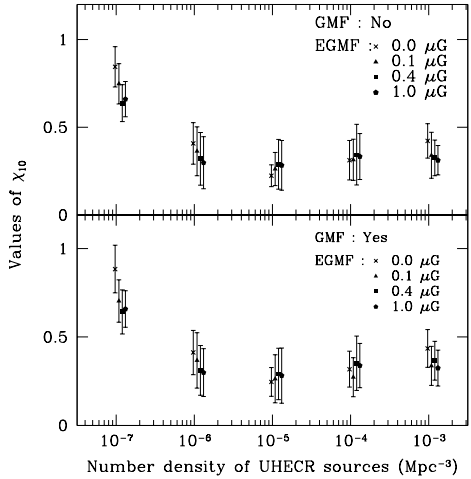}
\caption{$\chi_{10}$s as a function of the source number density 
in the normal source model. 
The error bars originate from 100 times source selection. 
The GMF is considered in the lower panel while not in the upper panel.}
\label{fig_chi10L0}
\end{center}
\end{figure}

\subsection{Constraints on the Source Number Density} \label{constraints}

The arrival distributions of UHE protons are constructed 
by a method explained in section \ref{method_prop}. 
Comparing the simulated arrival distribution with 
the observed arrival distribution statistically, 
we constrain the number density of UHECR sources. 

First, we try to restrict the number density 
from small-scale anisotropy. 
As mentioned above, 
the two-point correlation function is a statistical indicator 
of the small-scale anisotropy. 
Therefore, we compare the two-point correlation functions 
calculated from simulated arrival distributions 
and the functions calculated by AGASA result 
above $4 \times 10^{19}~{\rm eV}$ \cite{hayashida00}. 
In recent year, 
the authors of reference \cite{marco06a} pointed out 
that small-scale anisotropy by AGASA is statistically not consistent 
with the observed spectrum above $10^{20}$eV 
(above Greisen-Zatsepin-Kuz'min(GZK) cutoff \cite{greisen66,zatsepin66}) 
by them. 
Our source models also predict GZK cutoff 
even if our structured EGMF and GMF are taken into account, 
as shown in next subsection. 
Thus, we assume that the super-GZK events are of another origin. 
So, in this study we restrain the energy range of observed events we use 
up to $10^{20}$eV.

\begin{figure}
\begin{center}
\includegraphics[clip,width=\linewidth]{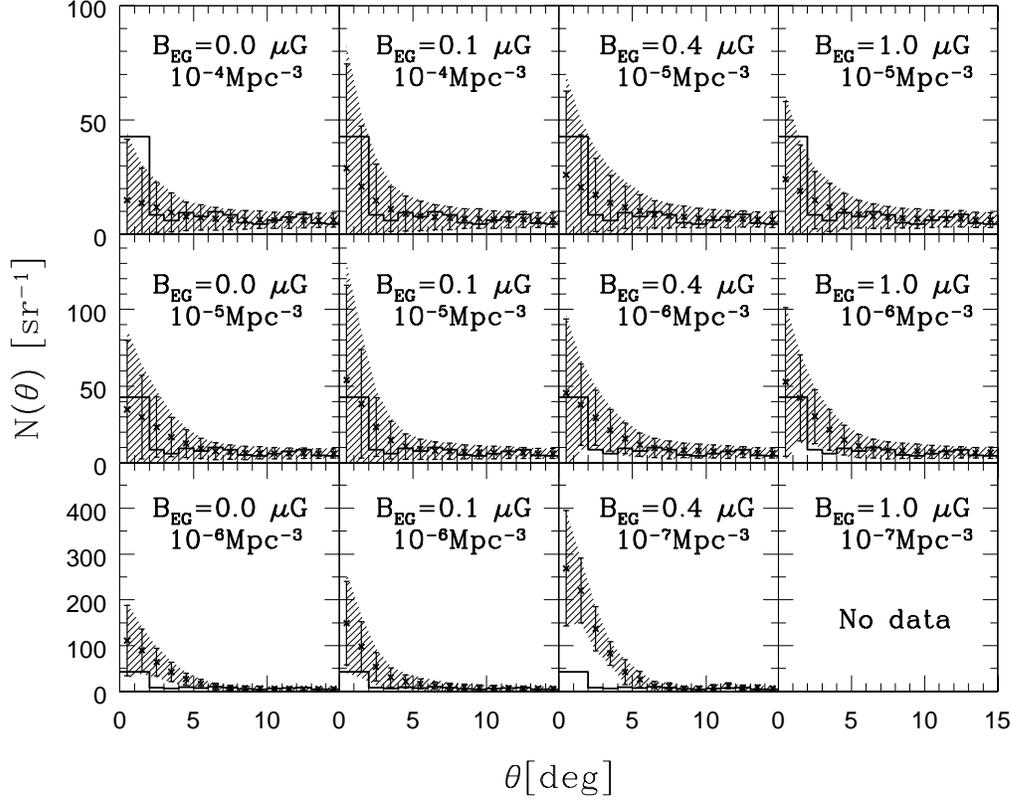}
\caption{The two-point correlation functions calculated 
from only source distributions that can predict the large-scale isotropy. 
The histograms are the observational result 
within $4 \times 10^{19} < E < 10^{20}~{\rm eV}$ (49 events) by AGASA. 
The error bars are from the event selection for finite events 
and the shaded regions show total 1$\sigma$ errors. 
The GMF is included.}
\label{fig_tp_isoL0}
\end{center}
\end{figure}

Figure \ref{fig_chi10L0} shows $\chi_{10}$ 
between two-point correlation functions calculated 
in the normal source model and AGASA data 
as a function of the number density of UHECR sources. 
The upper panel shows $\chi_{10}$s calculated without the GMF 
for $B=0.0\mu{\rm G}$ (crosses), $B=0.1\mu{\rm G}$ (triangles), 
$B=0.4\mu{\rm G}$ (squares), and $B=1.0\mu{\rm G}$ (pentagons) 
respectively. 
The error bars come from cosmic variance due to 100 source distributions 
for every number density. 
The lower panel shows the same, but GMF is included. 
The same number densities are represented 
despite four marks being a little shifted to the horizontal axis 
for easy visualization to see. 

The smaller $\chi_{10}$ gives the better number density. 
$\chi_{10}$s of $10^{-7}~{\rm Mpc^{-3}}$ are relatively larger 
than those of the others. 
Hence, $10^{-7}~{\rm Mpc^{-3}}$ cannot best reproduce the AGASA result. 
However, the remaining $\chi_{10}$s provide similar values. 
More discussion is needed in order to estimate 
the number density more precisely. 

UHECR source distribution must satisfy not only the small-scale anisotropy 
but also the large-scale isotropy. 
The large-scale isotropy is estimated quantitatively 
by harmonic amplitudes. 
Thus, we recalculate the two-point correlation function 
only from source distributions that predict 
the large-scale isotropy comparable to AGASA, 
which is quantified using the same method 
as in reference \cite{takami06}. 

Figure \ref{fig_tp_isoL0} shows such two-point correlation functions 
with several values of the number density and strengths of the EGMF. 
The GMF is also considered. 
The histograms are the two-point correlation functions 
calculated from AGASA events with energies from 
$4 \times 10^{19}$ to $10^{20}~{\rm eV}$ (49 events). 
The error bars are due to the finite number of event selection, and 
the shaded regions are total 1$\sigma$ error including 
errors due to different source distributions. 
The errors that originate from different source distributions are very small 
since source distributions that predicts large values of the function 
are excluded. 
Such source distributions cannot satisfy the large-scale isotropy.

We focus on the left three panels with $B_{\rm EG}=0.0 \mu{\rm G}$. 
In the upper panel, 
the mean values (points) are smaller than the observational data 
and cannot reproduce it within 1$\sigma$ error at small angular scale. 
On the other hand, 
source distributions with number densities of $10^{-6}~{\rm Mpc}^{-3}$ 
predict stronger anisotropy at the small angle scale 
even if the large-scale isotropy is satisfied. 
Thus, 
the appropriate number density that best reproduces 
the arrival distribution observed by AGASA is $\sim 10^{-5}~{\rm Mpc}^{-3}$ 
in the case of $B_{\rm EG}=0.0 \mu{\rm G}$. 
This is the same result as reference \cite{blasi04,kachelriess05} 
though they do not consider the GMF. 

\begin{figure}
\begin{center}
\includegraphics[clip,width=0.48\linewidth]{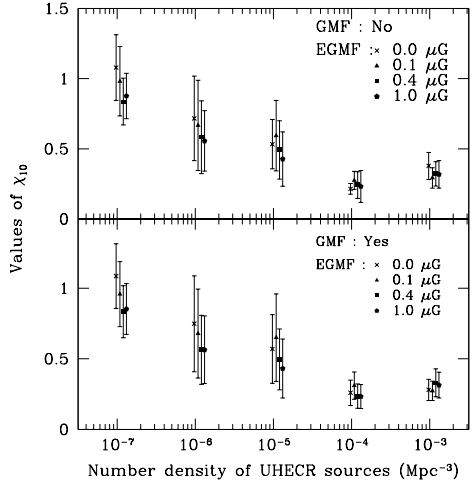}
\caption{Same as fig. \ref{fig_chi10L0}, 
but calculated in the luminosity-weighted source model.}
\label{fig_chi10L1}
\end{center}
\end{figure}

In the other panels, 
similar tendencies to the case of $B_{\rm EG}=0.0 \mu{\rm G}$ can be found. 
The EGMF diffuses cosmic rays during propagation, 
thereby weakening event clusterings. 
Therefore, smaller number densities ($\sim 10^{-6}~{\rm Mpc}^{-3}$) are 
favored in stronger EGMFs. 
However, in the case of $10^{-6}~{\rm Mpc}^{-3}$ 
with $B_{\rm EG} = 0.4, 1.0\mu {\rm G}$, 
number of source distributions used is only about 30\% 
of 100 source distributions 
while about 70\% at $10^{-5}~{\rm Mpc}^{-3}$. 
Although a stronger EGMF allows a smaller number density, 
the appropriate number density is close to $10^{-5}~{\rm Mpc}^{-3}$.

From the discussion above, 
the number density of $10^{-5}$ to $10^{-6}~{\rm Mpc}^{-3}$ 
can best reproduce the AGASA results, 
but, as you find in the upper panels, 
with uncertainty of about one order of magnitude. 
The result is almost unchanged even if the GMF is neglected.  

\begin{figure}
\begin{center}
\includegraphics[clip,width=\linewidth]{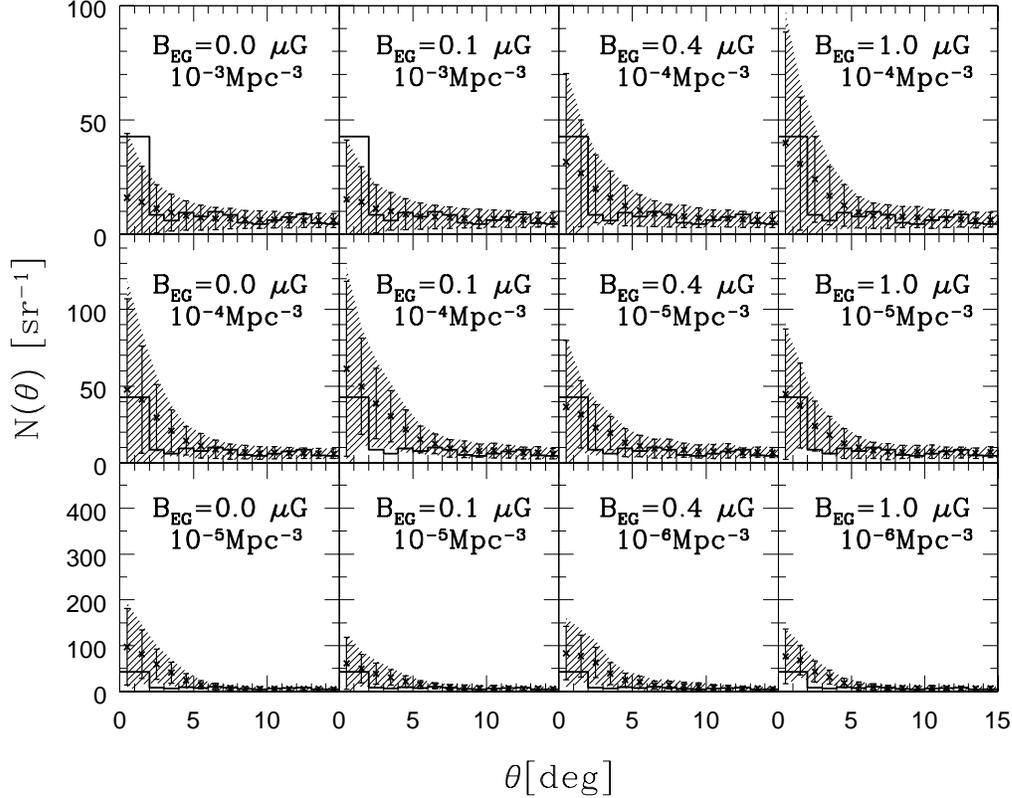}
\caption{Same as fig. \ref{fig_tp_isoL0}, 
but calculated in the luminosity-weighted source model.}
\label{fig_tp_isoL1}
\end{center}
\end{figure}

We can discuss similarly in the case of the luminosity weighted source model. 
Figure \ref{fig_chi10L1} shows the same as figure \ref{fig_chi10L0}, 
but calculated with the luminosity-weighted source model. 
The error bars are larger than those in figure \ref{fig_chi10L0} 
because of additional degree of freedom in the UHECR source luminosities. 
Mean values of $\chi_{10}$ are smallest at $10^{-4}~{\rm Mpc}^{-3}$, 
about one or two orders of magnitude larger than those 
in the case of the normal source model, 
but with some uncertainty. 
Thus, we discuss the arrival distributions 
that satisfy the large-scale isotropy again. 
Figure \ref{fig_tp_isoL1} shows the same as figure \ref{fig_tp_isoL0}, 
but calculated based on the luminosity-weighted source model. 
Two-point correlation functions predicted from the source number density 
of $\sim 10^{-3}~{\rm Mpc}^{-3}$ are smaller than the observational results 
at small scales. 
In the lower left panel, 
the predicted function is too large to reproduce the 
observed one at $2-4^{\circ}$ within 1$\sigma$ errors. 
When the strengths of the EGMF are stronger, 
values of the functions at small angles are more suppressed 
and the two-point correlation functions reproduce 
the observational result well. 
However, the number of source distributions with $10^{-5}~{\rm Mpc}^{-3}$, 
adopted in this calculation, is only about 10\% 
while the number with $10^{-4}~{\rm Mpc}^{-3}$ is about 50\%. 
Hence, the appropriate number density is thought to be closer to 
$10^{-4}~{\rm Mpc}^{-3}$.

As a result, in the luminosity-weighted source model, 
the number density of UHECR sources is $10^{-4}~{\rm Mpc}^{-3}$ 
with uncertainty. 
This is about one order of magnitude larger than that 
in the normal source model. 
As mentioned in section \ref{model}, 
the injection luminosity of sources is over 5 orders of magnitude 
in the luminosity-weighted source distribution. 
In these sources, 
more luminous sources can more strongly contribute to the arriving cosmic rays. 
Therefore, the number of such contributed sources 
seems smaller than the real number of cosmic ray sources. 

\subsection{Energy Spectra above $10^{19}~{\rm eV}$}

\begin{figure}[t]
\begin{center}
\includegraphics[width=0.48\linewidth]{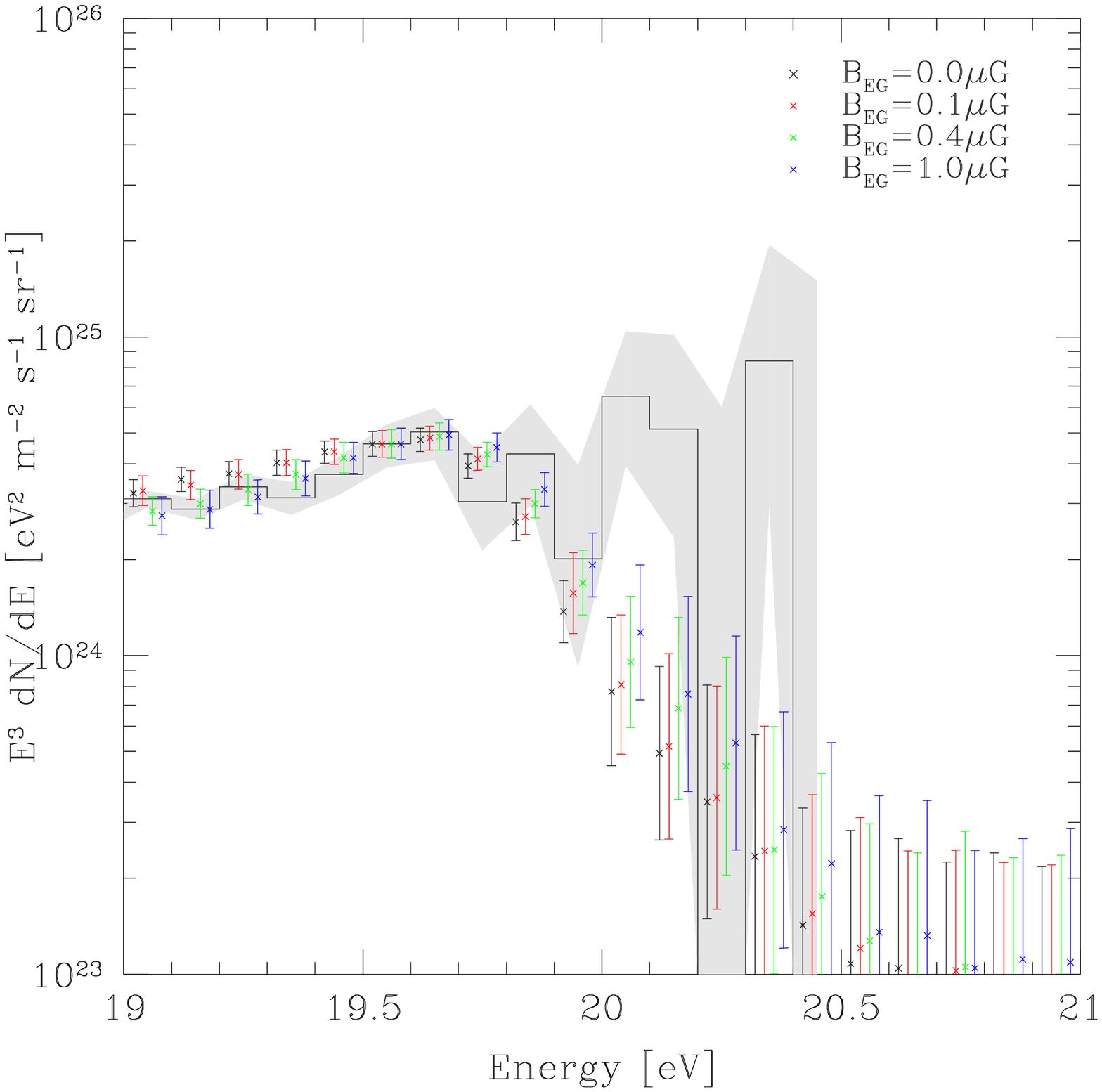} \hfill
\includegraphics[width=0.48\linewidth]{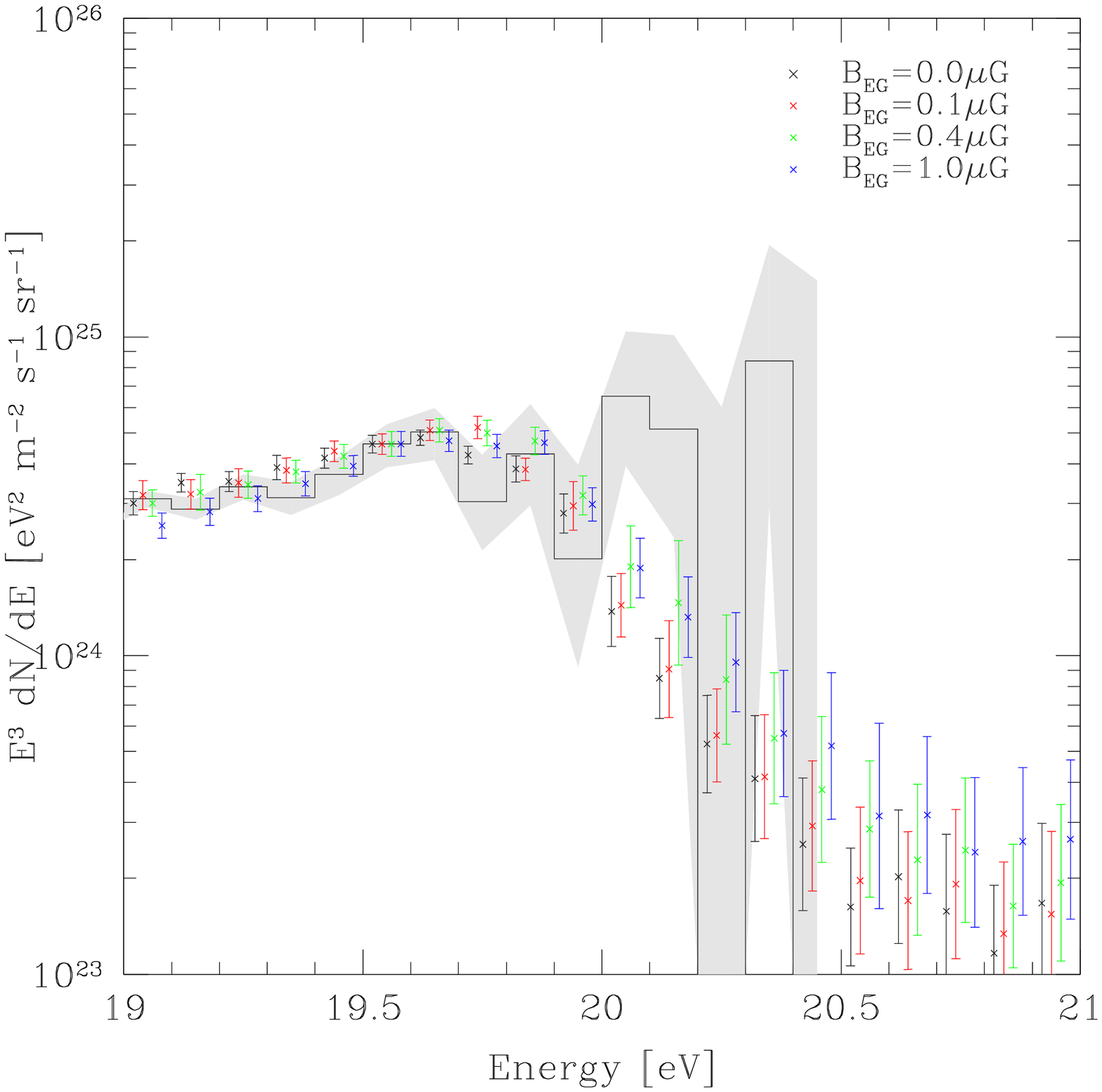}
\caption{Energy spectra of UHE protons calculated from 
all of 100 source distribution with $10^{-5}~{\rm Mpc}^{-3}$ 
in the normal source model ({\it left panel}), 
and from that with $10^{-4}~{\rm Mpc}^{-3}$ 
in the luminosity-weighted source model ({\it right panel}). 
Color points with error bars, which originates from 1$\sigma$ 
total statistical error, are calculated spectra 
with 0.0 ({\it black}), 0.1 ({\it red}), 0.4 ({\it green}), 
and 1.0 $\mu$G ({\it blue}). 
They are normalized at $10^{19.55}~{\rm eV}$. 
The histogram and shaded region are AGASA results\cite{takeda03}.}
\label{fig_spectrum}
\end{center}
\end{figure}

We discuss the dependency of the energy spectra on the magnetic field 
and source models. 
Figure \ref{fig_spectrum} shows the energy spectra 
calculated from 100 source distributions with 
$10^{-5}~{\rm Mpc}^{-3}$ in the normal source model ({\it left panel}), 
and $10^{-4}~{\rm Mpc}^{-3}$ 
in the luminosity-weighted source model ({\it right panel}). 
These spectra are predicted in the cases of 
$B_{\rm EG}$ = 0.0 ({\it black}), 0.1 ({\it red}), 0.4 ({\it green}), and 
1.0 $\mu$G ({\it blue}). 
The error bars are 1$\sigma$ total statistical 
error including cosmic variance. 
The GMF is also considered. 
The spectra are normalized at $10^{19.55}~{\rm eV}$. 
The histogram show the AGASA results \cite{takeda03} 
with shaded region representing statistical errors.

Around $10^{19}~{\rm eV}$, 
a stronger EGMF predicts a little lower flux 
since it lengthens the propagation distance 
and UHE protons lose their energies more readily. 
Under this normalization, 
a relatively higher flux is predicted at the highest energy region 
in a stronger EGMF. 
However, these are not statistically significant. 
Our structured EGMF cannot affect the energy spectrum. 
Our source models also cannot reproduce the super-GZK events. 
We assume that the super-GZK events are of other origins. 

\subsection{Future Prospects of our Approach}

In section \ref{constraints}, 
the number density of UHECR sources that can best reproduces 
the arrival distribution observed by AGASA is constrained. 
However, it has a large uncertainty that originates 
from the small number of observed events. 
Thus, our next interest is 
whether increase in observed events can determine 
the number density more accurately or not. 
In order to discuss this problem, 
it is necessary to compare our predicted arrival distribution 
to the arrival distribution observed by future observations. 
However, because we cannot know future results, 
we adopt an isotropic distribution 
of UHECR arrival directions 
as a template for future observations. 
A similar approach was adopted in reference \cite{marco06b}. 
However, simulation of \cite{marco06b} did not include magnetic fields. 
We discuss the future prospects taking into account GMF and EGMF. 

The small-scale anisotropy will be more strongly observed 
if UHECRs come from some point-like sources. 
Therefore, 
we compare the two-point correlation functions 
of our simulated arrival distribution to 
that calculated from 
the isotropic arrival distributions. 
To make the comparison, 
we introduce $\chi^2$ as 
\begin{equation}
\chi^2 \equiv \frac{1}{N_{\rm bin}} \sum_{\theta=1^{\circ}}
^{1^{\circ} \times N_{\rm bin}} 
\frac{\left[ N_{\rm sim}(\theta) - N_{\rm iso}(\theta) \right]^2}
{{\sigma_{\rm sim}(\theta)}^2 + {\sigma_{\rm iso}(\theta)}^2}, 
\end{equation}
where $N_{\rm sim}(\theta)$ and $N_{\rm iso}(\theta)$ are 
the two-point correlation functions calculated from simulated and 
isotropic arrival distributions respectively, 
and $\sigma_{\rm sim}(\theta)$ and  $\sigma_{\rm iso}(\theta)$ 
are 1$\sigma$ statistical errors 
of $N_{\rm sim}(\theta)$ and $N_{\rm iso}(\theta)$
due to the finite number of events. 
$N_{\rm bin}$ is set to be 5 to obtain information on 
the small-scale anisotropy. 
$\chi^2$s are distributed with their averages and variances 
for every source number density. 
If these distributions can be distinguished, 
we can argue that the source number density can be determined more accurately. 

\begin{figure}[t]
\begin{center}
\includegraphics[width=0.48\linewidth]{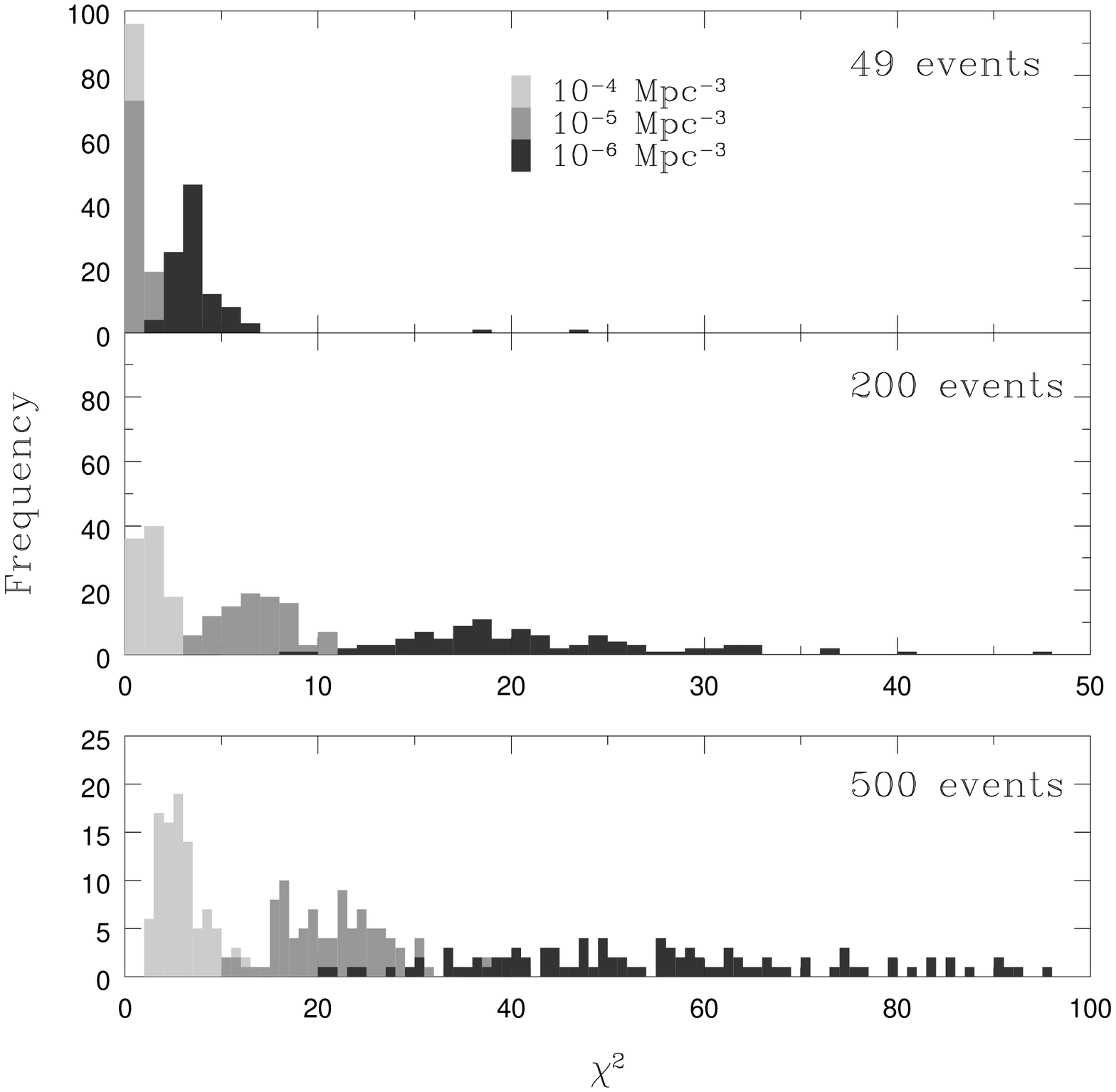} \hfill
\includegraphics[width=0.48\linewidth]{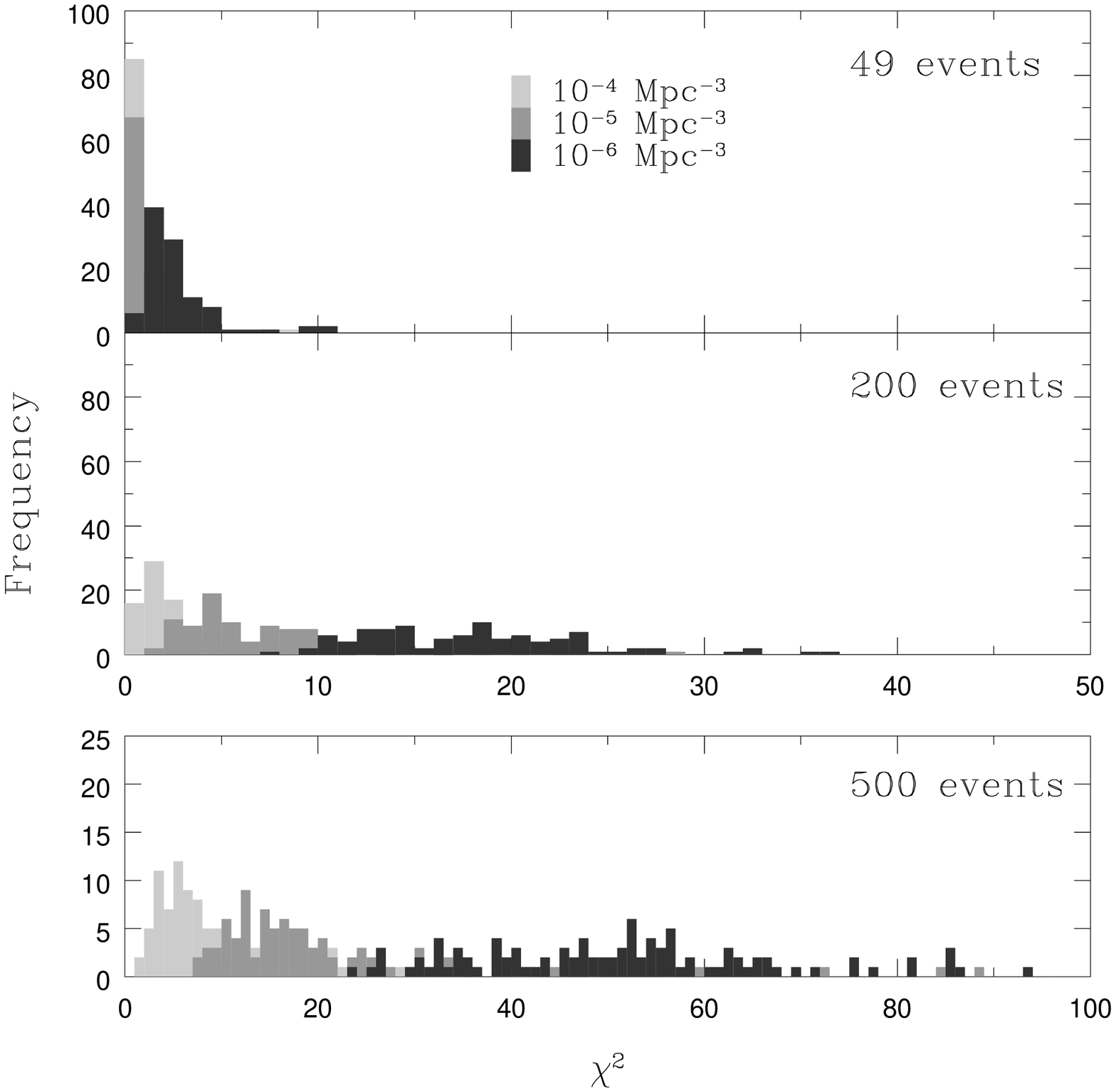}
\includegraphics[width=0.48\linewidth]{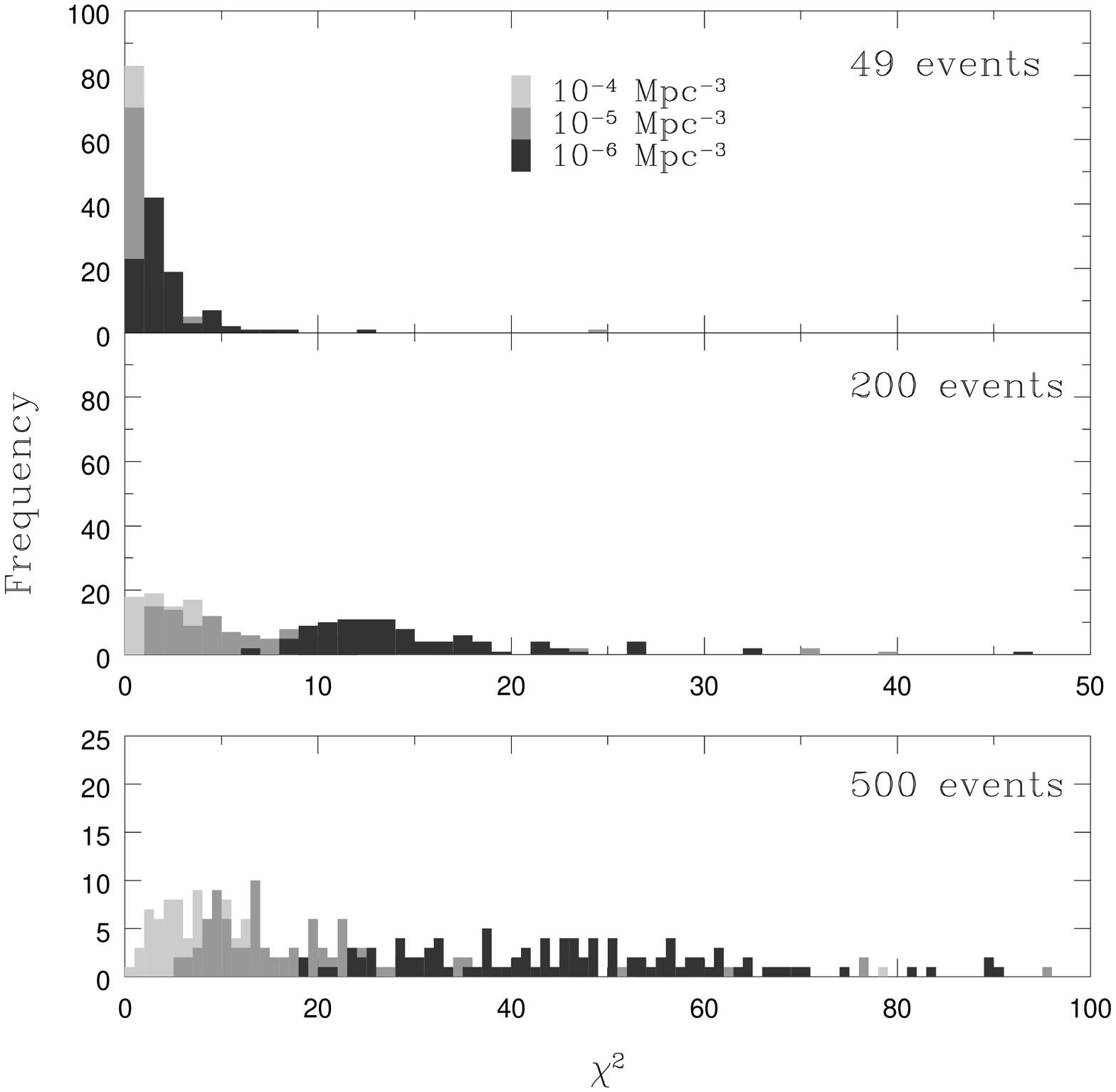} \hfill
\includegraphics[width=0.48\linewidth]{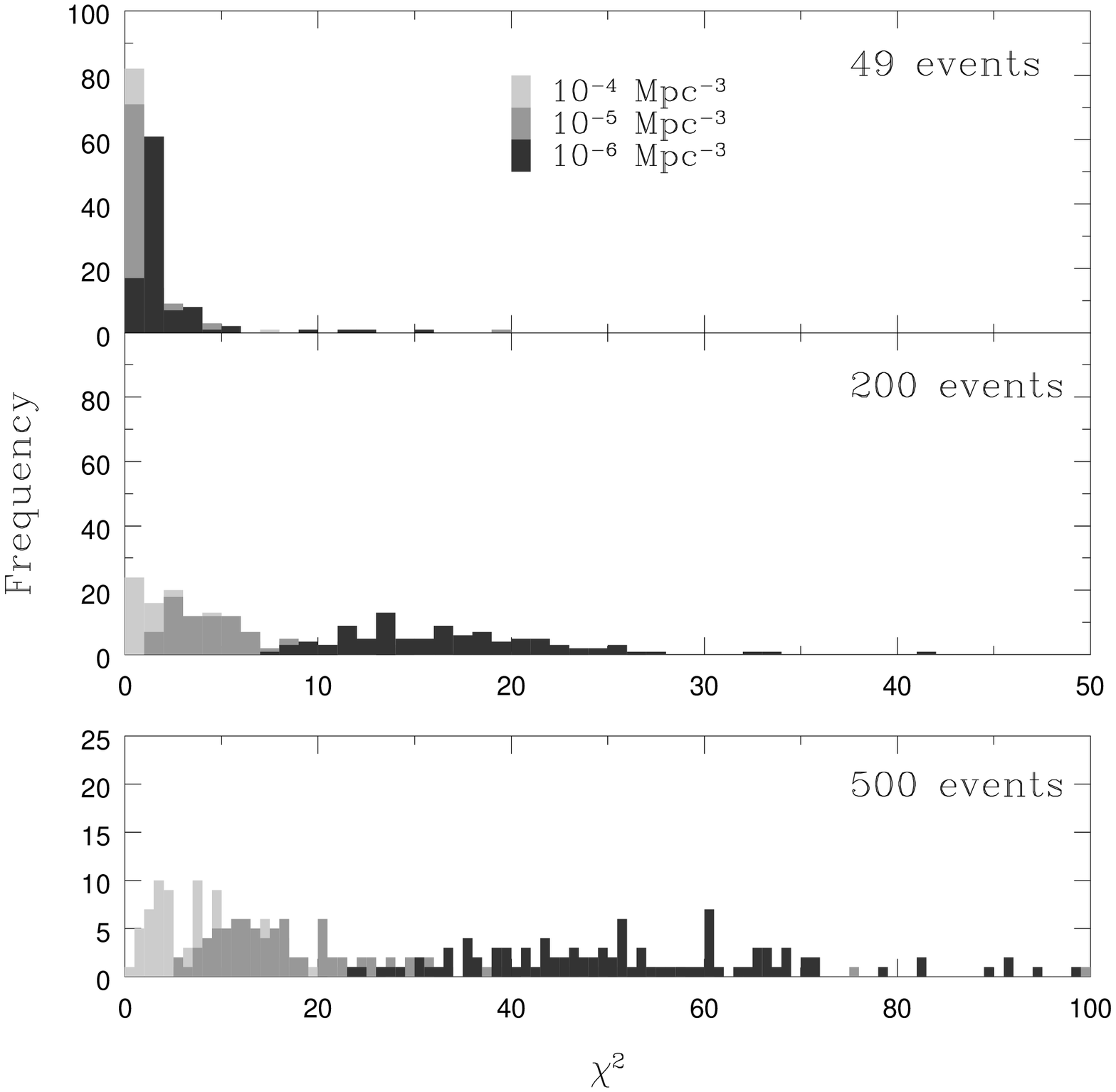}
\caption{Distributions of $\chi^2$s, 
calculated from arrival protons above $4 \times 10^{19}~{\rm eV}$, 
at several strengths 
of the EGMF in the normal source model. 
The GMF is not considered. 
The strengths of the EGMF are 
$0.0\mu \rm{G}$({\it upper left}), 
$0.1\mu \rm{G}$({\it upper right}), 
$0.4\mu \rm{G}$({\it lower left}), and 
$1.0\mu \rm{G}$({\it lower right}). 
The numbers of events are set to be 49 events 
within $-10^{\circ} < \delta < 80^{\circ}$, 
200 events and 500 events within the southern hemisphere 
to emulate Auger.}
\label{fig_chidisL0}
\end{center}
\end{figure}

Figure \ref{fig_chidisL0} shows distributions of $\chi^2$ 
in the normal source model 
at several EGMF strengths. 
The strengths of the EGMF are 0.0, 0.1, 0.4, and 1.0$\mu~{\rm G}$. 
These are calculated from arrival protons 
above $4 \times 10^{19}~{\rm eV}$. 
The numbers of events are set to 49 events 
within $-10^{\circ} < \delta < 80^{\circ}$ to emulate AGASA, and 
200 events and 500 events within $-90^{\circ} < \delta < 0^{\circ}$ 
to simulate Pierre Auger Observatory (Auger) \cite{augerhp}. 
The GMF is not considered, but 
the GMF hardly affects the result. 
The separation of the $\chi^2$ distribution from the first bin 
represents the appearance of difference from isotropic event distribution.

At current status (49 events), 
the $\chi^2$ distribution is localized at $\chi^2 = 1$ 
for $10^{-4}$ and $10^{-5}~{\rm Mpc}^{-3}$. 
This shows that 70-80\% of the source distributions with such number densities 
have an isotropy consistent with the large-scale isotropy reported by AGASA. 
The distributions with $10^{-6}~{\rm Mpc}^{-3}$ is a little shifted 
to the right for $B_{\rm EG} = 0.0\mu {\rm G}$. 
This shows that the arrival distribution expected 
for $10^{-5}~{\rm Mpc}^{-3}$ can be distinguished 
from that of $10^{-6}~{\rm Mpc}^{-3}$, to some extent. 
This is also supported in fig. \ref{fig_chi10L0}. 
The others with $10^{-6}~{\rm Mpc}^{-3}$ overlap 
the distribution for larger densities. 
Thus, the distinction between the number densities is difficult 
at present. 
It is remarkable that a stronger EGMF predicts a distribution 
biased to the left 
since it has the arrival distribution to be more isotropic. 

Next, we discuss a near future status with 200 events 
above $4 \times 10^{19}~{\rm eV}$, 
which corresponds to the number of events observed by Auger up to 2007. 
The distributions for $10^{-6}~{\rm Mpc}^{-3}$ do not 
overlap with those of the other number densities in all panels. 
Therefore, Auger can observe a difference 
between $10^{-6}~{\rm Mpc}^{-3}$ and more number densities 
while it is difficult to discriminate $10^{-5}~{\rm Mpc}^{-3}$ 
from $10^{-4}~{\rm Mpc}^{-3}$ 
except for the case of $B_{\rm EG} = 0.0\mu {\rm G}$. 
Source distributions with less than $10^{-5}~{\rm Mpc}^{-3}$ 
do not predict isotropy anymore. 

Accumulation of 500 events gives rise to larger separation. 
Even the arrival distributions expected from $10^{-4}~{\rm Mpc}^{-3}$ 
are distinguishable from isotropic distribution. 
For $B_{\rm EG} = 0.0\mu {\rm G}$, 
the three distributions can be completely separate, 
and, we can determine the source number density 
from an observational arrival distribution. 
However, 
when the EGMF is considered, 
the distributions with $10^{-5}~{\rm Mpc}^{-3}$ 
overlap with those of $10^{-4}~{\rm Mpc}^{-3}$, 
since the stronger EGMF diffuses the UHECR arrival directions 
and their arrival distribution becomes more isotropic. 
More observation is required. 

\begin{figure}[t]
\begin{center}
\includegraphics[width=0.48\linewidth]{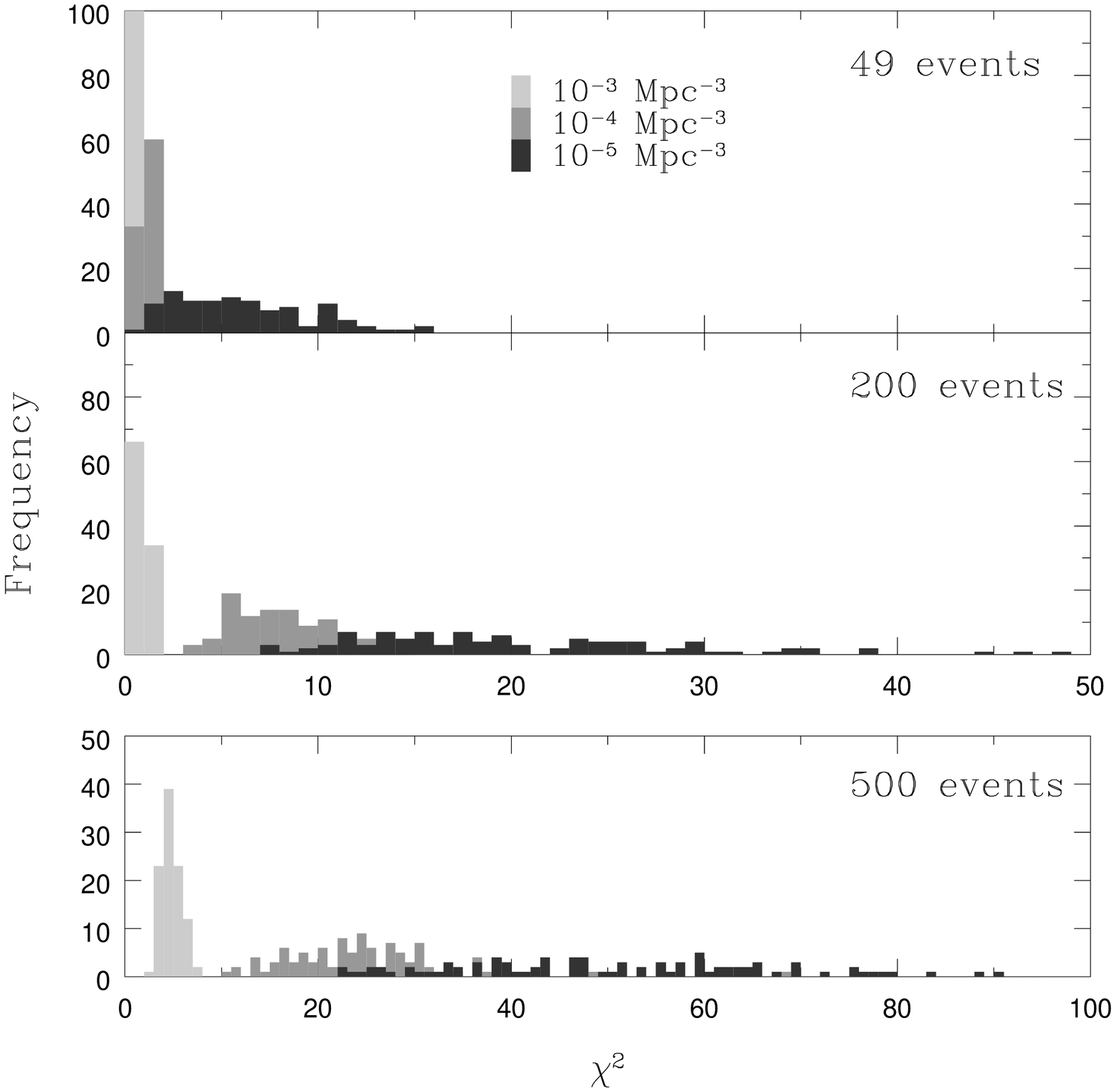} \hfill
\includegraphics[width=0.48\linewidth]{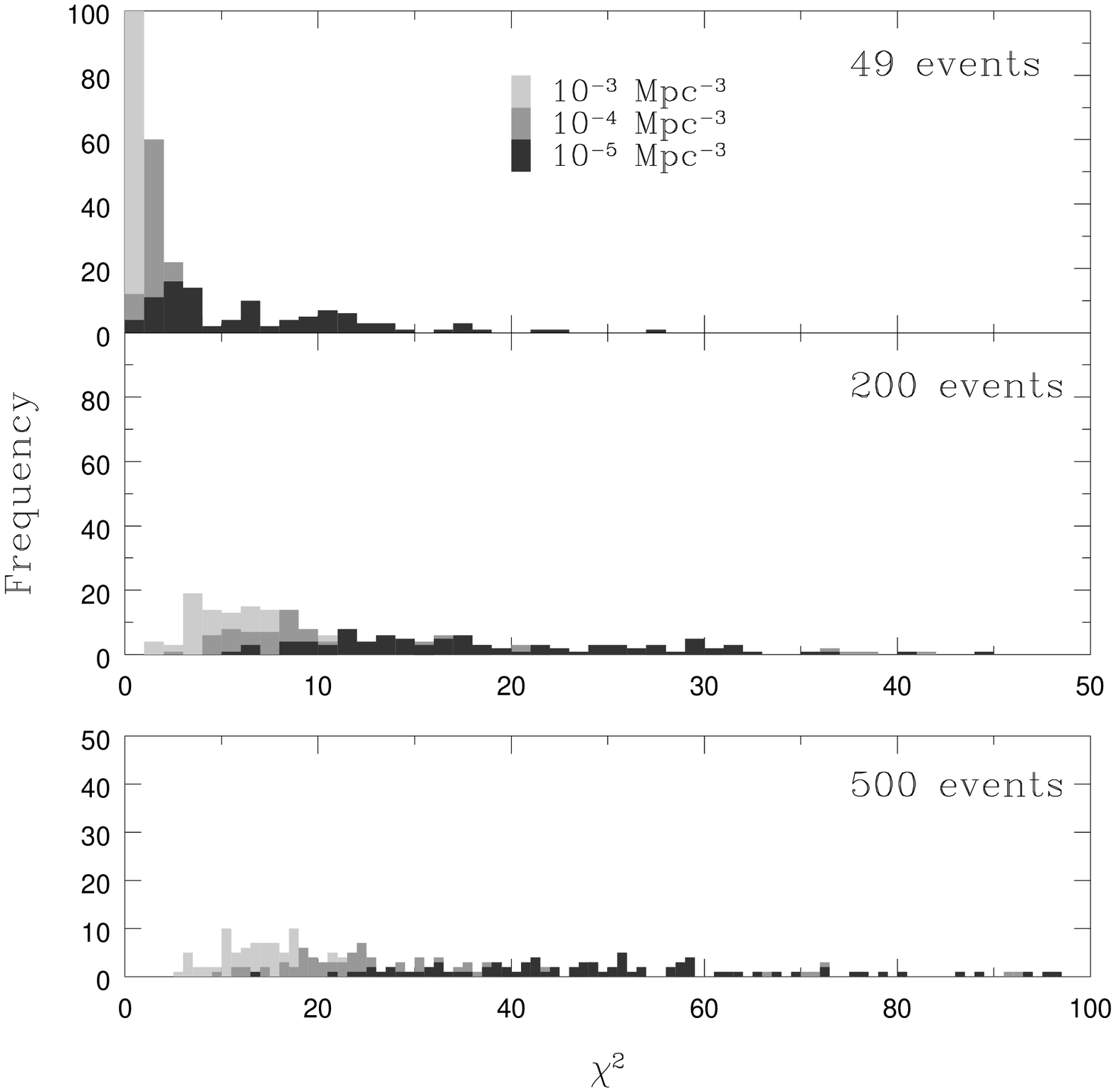}
\includegraphics[width=0.48\linewidth]{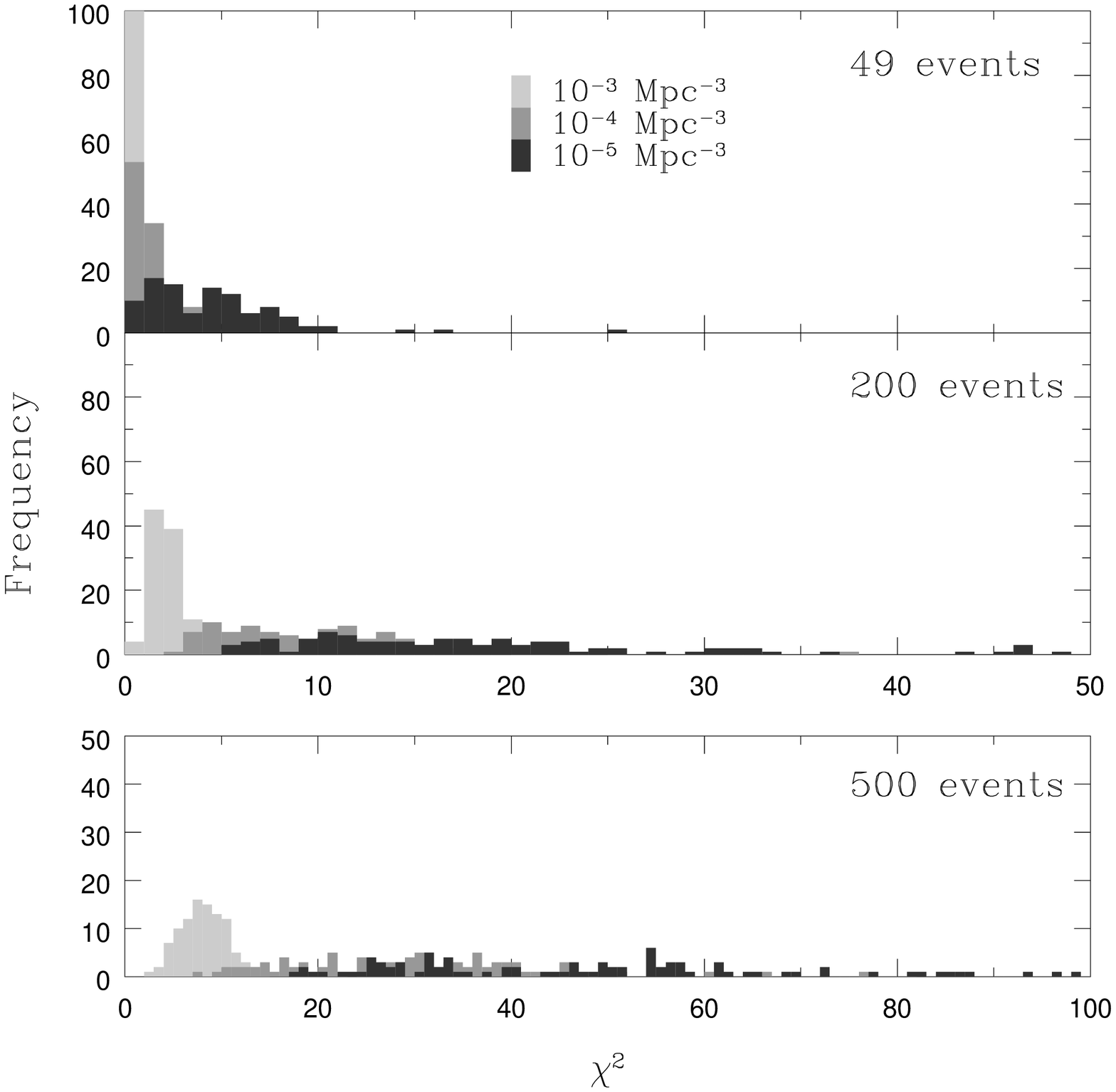} \hfill
\includegraphics[width=0.48\linewidth]{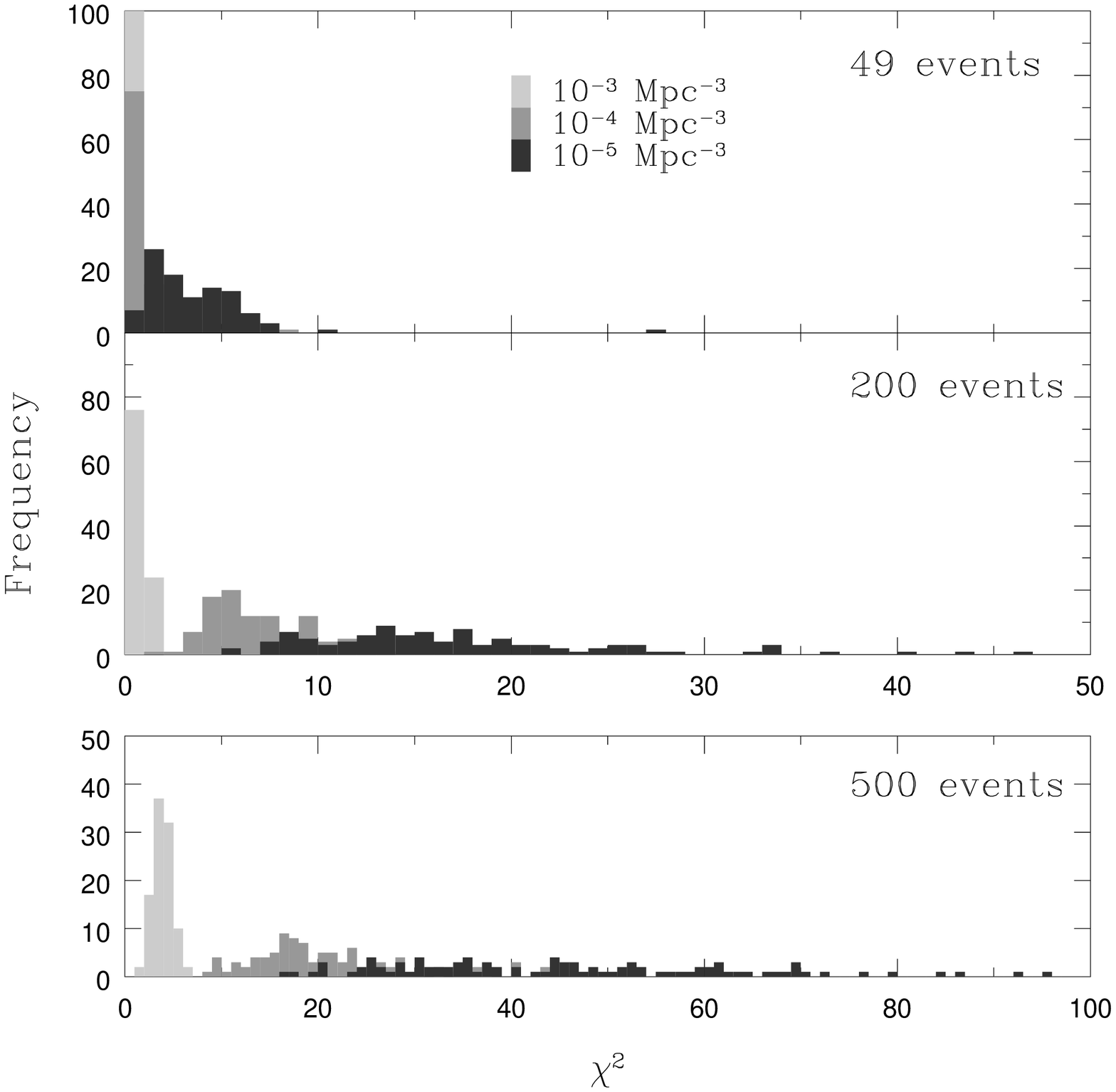}
\caption{Same as fig.\ref{fig_chidisL0}, 
but the luminosity-weighted source model is adopted.}
\label{fig_chidisL1}
\end{center}
\end{figure}

In the luminosity-weighted source model, 
we can discuss as above for $10^{-3},~10^{-4}$, and $10^{-5}~{\rm Mpc}$. 
The graph is shown in figure \ref{fig_chidisL1}.
At present (49 events), 
$10^{-5}~{\rm Mpc}^{-3}$ has distributions with large width. 
This fact reflects that only 10\% of the source distributions 
can reproduce the large-scale isotropy. 
For 200 event observation, 
the three distributions are separated more largely. 
In this case, one question occurs. 
The source distributions with $10^{-3}~{\rm Mpc}^{-3}$ in 
$B_{\rm EG} = 0.1\mu {\rm G}$ do not predict an isotropy 
similar to that in $B_{\rm EG} = 0.0\mu {\rm G}$. 
This seems to be strange. 
This is because the arriving cosmic-rays from near sources 
relatively increase at highest energies by the EGMF. 
Source distributions with $10^{-3}~{\rm Mpc}^{-3}$ 
contain the most number of UHECR sources. 
Because near sources cause anisotropy, 
isotropy is worse than that in $B_{\rm EG} = 0.0\mu {\rm G}$. 
When the EGMF is stronger, 
the diffusion of cosmic-rays can overcome the anisotropy. 
Hence, the distributions of $\chi^2$ for $B_{\rm EG} = 0.4$ and 
$1.0\mu {\rm G}$ are shifted to the left. 
200 event observation enables us to distinguish $10^{-3}~{\rm Mpc}^{-3}$ 
with $10^{-4}~{\rm Mpc}^{-3}$. 
The arrival distribution from less than $10^{-4}~{\rm Mpc}^{-3}$ 
is distinguishable from the isotropic distribution. 

For 500 event observation, 
$10^{-3}~{\rm Mpc}^{-3}$ is separated from the others 
because of strong deflection providing UHECR arrival distributions isotropy. 
However, the distributions of 
$10^{-4}~{\rm Mpc}^{-3}$ and $10^{-5}~{\rm Mpc}^{-3}$ 
have broad width 
due to the degree of freedom of the source luminosity. 
Even 500 events observation cannot enable us to discriminate 
these source number densities, 
but to discriminate $10^{-3}~{\rm Mpc}^{-3}$ from the isotropic distribution.

%%%%%%%%%%%%%%%%%%%%%%%%%%%%%%%%%%%%%%%%%
%%%%%%%%%%%%%%%%%%%%%%%%%%%%%%%%%%%%%%%%%
\section{Summary \& Discussion} \label{summary}
%%%%%%%%%%%%%%%%%%%%%%%%%%%%%%%%%%%%%%%%%
%%%%%%%%%%%%%%%%%%%%%%%%%%%%%%%%%%%%%%%%%

In this study, 
we discuss the possibility of accurately estimating 
the source number density of UHECRs with small-scale anisotropy. 
Comparison between simulated arrival distribution and 
the observational results enables us to 
estimate the source number density. 
In order to construct the arrival distribution, 
we calculate the propagation of UHE protons 
in a structured EGMF with several strengths 
consistent with measurements of magnetic field 
in clusters of galaxies. 
The GMF is also considered. 
We find that the source number density of  
$10^{-5}~{\rm Mpc}^{-3}$ in the normal source model and 
$10^{-4}~{\rm Mpc}^{-3}$ in the luminosity-weighted source model 
can best reproduce the AGASA results, 
which are weakly dependent on strength of our structured EGMF. 
However, these have large uncertainty of about one order of magnitude 
due to the small number of observed events. 

So, we discuss the possibility that future observations decrease 
the uncertainty. 
In the normal source model, 
Auger can distinguish $10^{-5}~{\rm Mpc}^{-3}$ and $10^{-6}~{\rm Mpc}^{-3}$ 
sufficiently by our method in the near future. 
If the structured EGMF is zero or very weak, 
$10^{-4}~{\rm Mpc}^{-3}$ is also discriminated from the less number density 
in 500 event observation above $4 \times 10^{19}~{\rm eV}$. 
In stronger EGMF, more observations are requested 
because cosmic rays are diffused more strongly. 
Number of events that needed for the distinction depends on EGMF strength. 
In the luminosity-weighted source model, 
$10^{-3}~{\rm Mpc}^{-3}$ can be distinguished from the less number density 
by Auger. 
The distinction between the less number densities is difficult 
due to large uncertainty which originates 
from different injection powers of the sources. 

\begin{figure}[t]
\begin{center}
\includegraphics[width=0.48\linewidth]{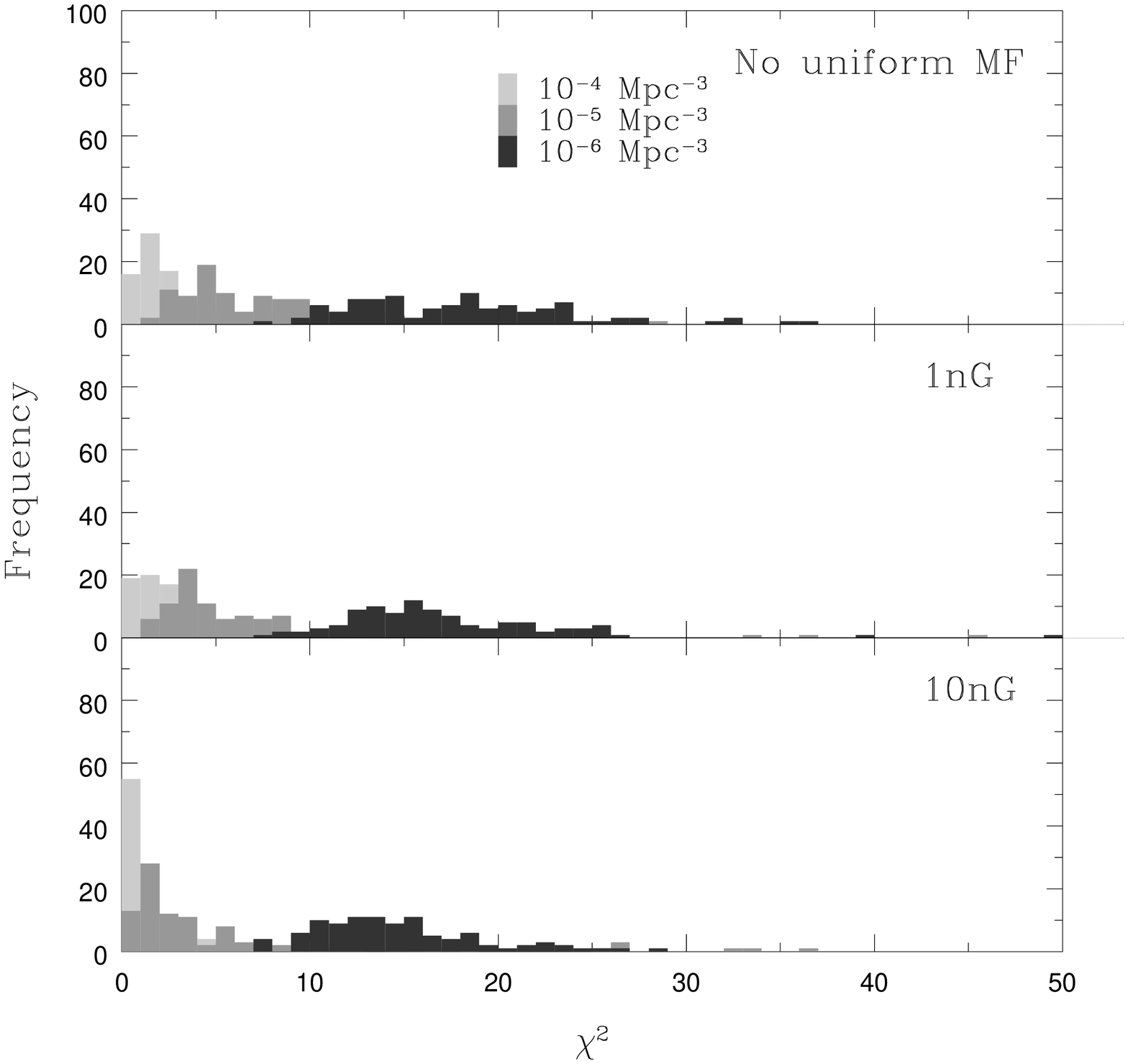} \hfill
\includegraphics[width=0.48\linewidth]{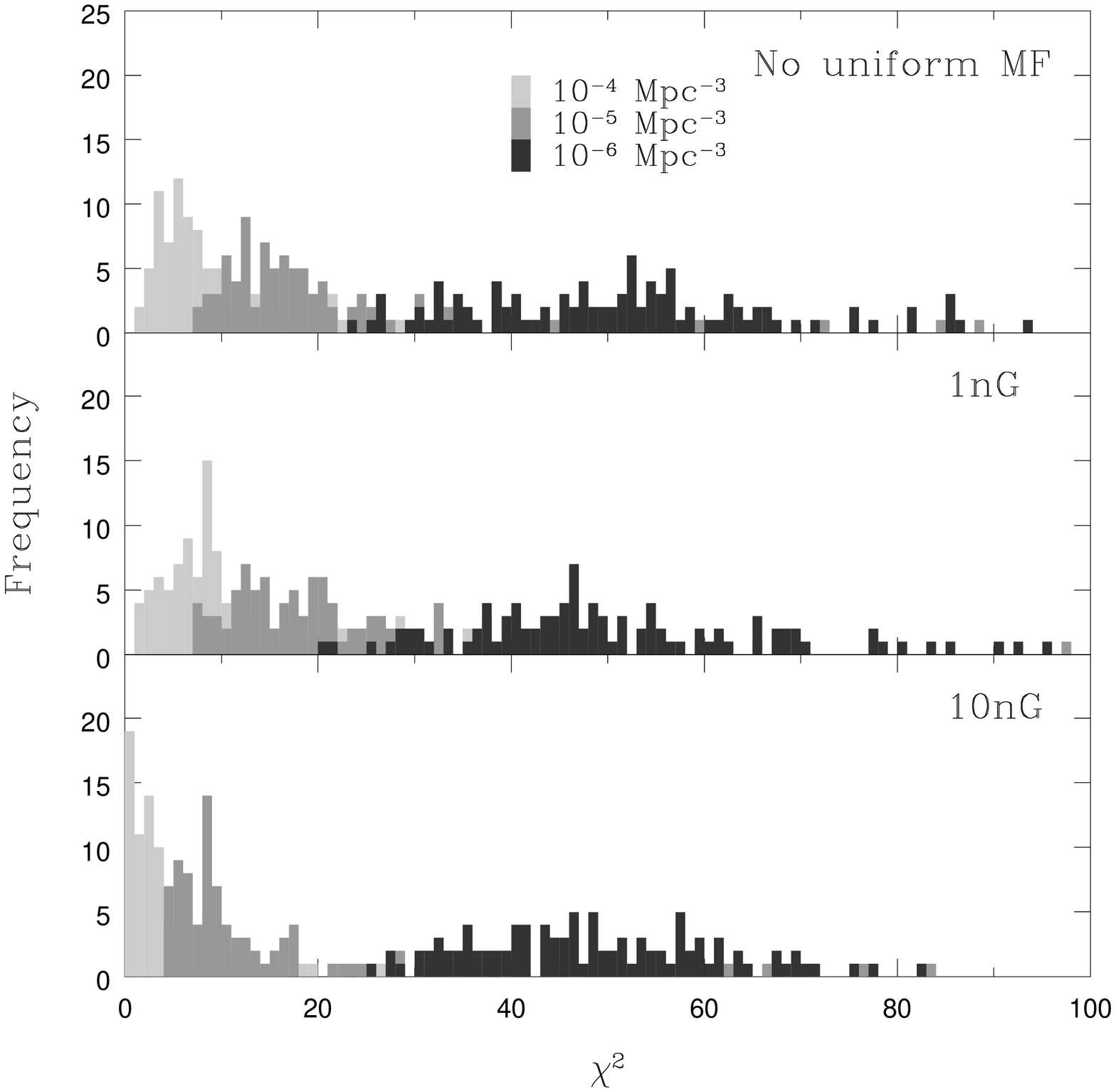}
\caption{Distribution of $\chi^2$s calculated 
from arrival protons above $4 \times 10^{19}~{\rm eV}$, 
in the normal source model. 
The GMF is not considered 
The strength of the structured EGMF is normalized to 0.1 $\mu$G 
and that of an uniform turbulent field is 0 ({\it upper panel}), 
1 nG ({\it middle panel}), and 10 nG ({\it lower panel}). 
The numbers of events are set to be 200 ({\it left}) 
and 500 events ({\it right}) within the southern hemisphere.}
\label{fig_chi_offset}
\end{center}
\end{figure}

We exclusively adopt the AGASA results in this study. 
However, High Resolution Fly's Eye (HiRes) claims 
no significant small-scale anisotropy, contrary to AGASA \cite{abbasi04}. 
This discrepancy is one of well-known problems in UHECR experiments. 
At present, 
this is not statistically significant due to the small number 
of observed events \cite{yoshiguchi04}. 
It will be able to solved by new experiments with large aperture, 
such as Auger, Telescope Array \cite{tahp}, 
and Extreme Universe Space Observatory \cite{eusohp}. 
Hence, it is possible that these experiments do not observe 
sufficient small-scale anisotropy in the future. 
If Auger does not observe small-scale clusterings during 2007 
(maybe it detects about 200 events above $4 \times 10^{19}~{\rm eV}$), 
the source number density is estimated at about $10^{-4}~{\rm Mpc}^{-3}$ 
or more in the normal source model by definition of $\chi^2$ 
in figure \ref{fig_chidisL0}. 
It is comparable with number density of active galactic nuclei\cite{loveday92}.

Our EGMF model within 100 Mpc has about 95\% of volume without magnetic field. 
In our model, uniform magnetic field is not considered. 
According to an upper limit mentioned in section \ref{introduction}, 
deflection angle of UHE protons with energy of $E$ during 
propagation of distance, $d$, is estimated as 
\begin{equation}
\theta < 3^{\circ} \left( \frac{E}{10^{20}~{\rm eV}} \right)^{-1} 
\left( \frac{d}{100~{\rm Mpc}} \right)^{1/2} 
\left( \frac{l_c}{1~{\rm Mpc}} \right)^{1/2} 
\left( \frac{B}{1~{\rm nG}} \right). 
\end{equation}
This deflection is expected to generate 
more isotropic arrival distribution of UHECRs. 
Therefore, uniform turbulent magnetic field affects determination 
of the source number density. 

As a demonstration, 
we show $\chi^2$ distribution for $B_{\rm EG} = 0.1~\mu{\rm G}$, 
including the uniform field, in figure \ref{fig_chi_offset}. 
The normal source model is adopted. 
The number of events is set to be 200 ({\it left panel}), 
and 500 ({\it right panel}). 
The strengths of the uniform turbulent field are 0 ({\it upper panel}), 
1 ({\it middle panel}), and 10 nG ({\it lower panel}). 
The distributions are shifted to lower value in stronger uniform 
turbulent field. 
The shifts are larger in more source number density. 
In stronger turbulent uniform magnetic field, 
number of events needed for the distinction between $10^{-5}$ and 
$10^{-6}~{\rm Mpc}^{-3}$ is smaller while 
it becomes difficult to discriminate $10^{-4}$ from $10^{-5}~{\rm Mpc}^{-3}$ 
due to the diffusion of cosmic rays. 
Strength of uniform EGMF is also 
important for estimating the source number density. 

In this work, 
we adopt isotropic arrival distribution of UHECRs as a template 
of a future result 
since results of new experiments are still unpublished. 
Auger should detect a few times more number of UHE events 
than that of AGASA 
since it started observation about three years ago. 
Its result will provide us beneficial information 
on the nature of UHECR sources.

\subsubsection*{Acknowledgements:} 
The work of H. T. is supported by Grants-in-Aid from JSPS Fellows. 
The work of K. S. is supported by Grants-in-Aid for 
Scientific Research provided by the Ministry of Education, Science 
and Culture of Japan through Research Grants 
S19104006.


\begin{thebibliography}{}
\bibitem{takeda99} M. Takeda et al., Astrophys.~J. 522 (1999) 225.
\bibitem{takeda01} M. Takeda et al., Proc. 27th Int. Cosmic Ray Conf. (Hamburg)  1 (2001) 341
\bibitem{kronberg94} P.P. Kronberg, Rep.~Prog.~Phys. 57 (1994) 325
\bibitem{yoshiguchi03} H. Yoshiguchi et al., Astrophys.~J. 586 (2003) 1211 (erratum 601 (2004) 592)
\bibitem{aloisio04} R. Aloisio, V. Berezinsky, Astrophys.~J. 612 (2004) 900
\bibitem{berezinsky06} V. Berezinsky, A. Gazizov, S. Grigorieva, Phys.~Rev.~D 74 (2006) 043005
\bibitem{vallee04} J.P. Vall$\rm{\acute{e}}$e, New Astron.Rev. 48 (2004) 763
\bibitem{sigl03} G. Sigl, F. Miniati, T.A. Ensslin, Phys.~Rev.~D 68 (2003) 043002
\bibitem{sigl04} G. Sigl, F. Miniati, T.A. Ensslin, Phys.~Rev.~D 70 (2004) 043007
\bibitem{dolag05} K. Dolag et al., JCAP 0501 (2005) 009
\bibitem{takami06} H. Takami, H. Yoshiguchi, K. Sato, Astrophys.~J. 639 (2006) 803 (erratum 653 (2006) 1584)
\bibitem{saunders00} W. Saunders et al., MNRAS 317 (2000) 55
\bibitem{takeuchi03} T.T. Takeuchi, K. Yoshikawa, T.T. Ishii, Astrophys.~J. 587 (2003) L89 (erratum 606 (2004)  L171)
\bibitem{alvarez02} J. Alvarez-Muniz, R. Engel, T. Stanev, Astrophys.~J. 572 (2002) 185
\bibitem{yoshiguchi04a} H. Yoshiguchi, S. Nagataki, K. Sato, Astrophys.~J. 607 (2004) 840
\bibitem{berezinsky88} V. Berezinsky, S.I. Grigorieva, A\&A 199 (1988) 1
\bibitem{yoshida93} S. Yoshida, M. Teshima, Prog. Theor. Phys. 89 (1993) 833
\bibitem{chodorowski92} M.J. Chodorowski, A.A. Zdziarske, M. Sikora, Astrophys.~J. 400 (1992) 181
\bibitem{mucke00} A. Mucke, et al., Comput.~Phys.~Commun. 124 (2000) 290
\bibitem{marco03} D. de Marco, P. Blasi, A.V. Olinto, Astropart.~Phys. 20 (2003) 53
\bibitem{berezinsky05} V. Berezinsky, A.Z. Gazizov, S.I. Grigorieva, Phys.~Lett. B612 (2005) 147
\bibitem{hayashida00} N. Hayashida et al., Astron.~J. 120 (2000) 2190
\bibitem{marco06a} D. de Marco, P. Blasi, and A.V. Olinto, JCAP 0601 (2006) 002
\bibitem{greisen66} K. Greisen, Phys.~Rev.~Lett. 16 (1966) 748
\bibitem{zatsepin66} G.T. Zatsepin, V.A. Kuz'min, JETP Lett. 4 (1966) 78
\bibitem{blasi04} P. Blasi, D. de Marco, Astropart.~Phys. 20 (2004) 559
\bibitem{kachelriess05} M. Kachelriess, D. Semikoz, Astropart.~Phys. 23 (2005) 486
\bibitem{takeda03} M. Takeda et al., Astropart.~Phys. 19 (2003) 447
\bibitem{marco06b} D. de Marco, P. Blasi, and A.V. Olinto, JCAP 0607 (2006) 015
\bibitem{augerhp} http://www.auger.org
\bibitem{abbasi04} R.U. Abbasi et al., Astrophys.~J. 610 (2004) L73
\bibitem{yoshiguchi04} H. Yoshiguchi, S. Nagataki, K. Sato, Astrophys.~J. 614 (2004) 43
\bibitem{tahp} http://www-ta.icrr.u-tokyo.ac.jp/index\_en.html
\bibitem{eusohp} http://euso.riken.go.jp
\bibitem{loveday92} J. Loveday et al., Astrophys.~J. 390 (1992) 338
\end{thebibliography}
\end{document}